\documentclass[intlimits,twoside,a4paper]{article}

\usepackage[cp1251]{inputenc}
\usepackage[eqsecnum]{cmpj3}
\usepackage{bm}
\usepackage{array}

\issue{2021}{24}{2}{23702}
\doinumber{10.5488/CMP.24.23702}
\title[Electronic structure and elastic properties of Cd${}_{16}$Se${}_{15}$Te]%
{Electronic structure and elastic properties of Cd${}_{16}$Se${}_{15}$Te solid state solution: first principles study}

\author[A.~I. Kashuba \textsl{et al.}]{A.~I. Kashuba\refaddr{label1},
        B. Andriyevsky\refaddr{label2},
        H.~A. Ilchuk\refaddr{label1},
        R.~Yu. Petrus\refaddr{label1},
        T.~S. Malyi\refaddr{label3},
        I.~V.~Semkiv\refaddr{label1}}
\addresses{
\addr{label1} Lviv Polytechnic National University, 12 Bandera Str., 79013 Lviv, Ukraine
\addr{label2} Koszalin University of Technology, 2 Sniadeckich Str., 75-453 Koszalin, Poland
\addr{label3} Ivan Franko National University of Lviv, 8 Kyryla and Mefodiya Str., 79005 Lviv, Ukraine
}
\date{Received July 27, 2020, in final form March 8, 2021}
\begin{document}

\maketitle

\begin{abstract}
The electronic band structure and elastic properties of the Cd${}_{16}$Se${}_{15}$Te solid state solution in the framework of the density functional theory calculations are investigated. The structure of the sample is constructed on the original binary compound CdSe, which crystallizes in the cubic phase. Based on the electronic band structure, the effective mass of electron, heavy hole, light hole, spin-orbit effective masses and reduced mass in G point are calculated. In addition, the exciton binding energy, refractive index and high-frequency dielectric constant are calculated. The Young modulus, shear modulus, bulk modulus and Poisson ratio are calculated theoretically. Based on the results of elastic coefficients, the value of acoustic velocity and Debye temperature is obtained.
\end{abstract}

\section{Introduction}

The A${}^{\text{II}}$B${}^{\text{VI}}$ semiconductor compounds and their solid solutions, such as CdSeTe, have important applications such as infrared detectors, solar cells and other devices~\cite{1, 2, 3, 4, 5, 6, 7, 8, 9, 10}. Despite the recent intensive experimental and theoretical studies of these materials some of the fundamental parameters are currently unknown, primarily mechanical properties. This information is very important for modelling and building certain parts for optical and electronic devices. Some information regarding the physical properties of CdSeTe solid state solutions is presented in the previous publications~\cite{9, 10, 11, 12, 13, 14, 15, 16}. Most of them are focused on the study of fundamental optical and electrical parameters.

The CdSeTe solid state solutions may be considered as a particular class of semiconductor materials having promising optical parameters for practical applications. This is due to the clear band gap $\textit{E}_{g}$ dependence on the selenium content $\textit{x}$ of the material, CdTe${}_{1-\textit{x}}$Se${}_{\textit{x}}$. The band gap changes from $\textit{E}_{g}=$~1.44~eV for CdTe~\cite{17, 18, 19, 20}) to $\textit{E}_{g}=$~1.68~eV for CdSe~\cite{21, 22, 23}. It is also important  that the above values of band gaps correspond to the near IR photon energy range, in which many commercial laser sources are accessible. The CdTe compounds have a cubic (zinc blende) structure~\cite{24, 25}, whereas CdSe compounds, depending on the growth conditions, may have both zinc blende and wurtzite (hexagonal) structures at normal conditions~\cite{22, 26, 27}. According to the phase diagram~\cite{28}, crystallization of CdTe${}_{1-\textit{x}}$Se${}_{\textit{x}}$ is possible into a cubic structure with selenium concentration $\textit{x}< 0.33$ and into wurtzite structure with the corresponding concentration $\textit{x}> 0.55$.

Among numerous theoretical studies of the electron band structure~\cite{3, 7, 8, 14, 16, 20, 22, 25, 29, 30}, optical parameters~\cite{14, 15, 22, 25, 30} and effective mass~\cite{30, 31, 32} of CdSe-CdTe compounds, we have not found the corresponding results for the effective mass and elastic properties of Cd${}_{16}$Se${}_{15}$Te  solid state solution corresponding to the tellurium content $\textit{x}= 0.0625$. This value of tellurium content may be characteristic of the CdSe${}_{1-\textit{x}}$Te${}_{\textit{x}}$ solid state solutions possessing the zinc blend cubic structure.

In this paper, we report on electronic and elastic properties of the Cd${}_{16}$Se${}_{15}$Te solid state solution. The structure of cubic crystal CdSe (Cd${}_{16}$Se${}_{16}$) is used as the initial crystal structure of Cd${}_{16}$Se${}_{15}$Te solid state solution, where one selenium atom is changed by the tellurium atom. Results of the band structure calculations of Cd${}_{16}$Se${}_{15}$Te are used to obtain the effective mass, exciton binding energy, refractive index and high-frequency dielectric constant. The acoustic velocity and Debye temperature of the sample is calculated using the calculated elastic modules.

\section{Methods of calculation}

The band structure and the related properties of Cd${}_{16}$Se${}_{15}$Te solid state solutions were calculated in the framework of the density functional theory (DFT)~\cite{33} using CASTEP code~\cite{34}. In the present calculations, the generalized gradient approximation (GGA) and the Perdew-Burke-Ernzerhof (PBESOL) exchange-and-correlation functional~\cite{35} were utilized. The interaction of electrons with atomic cores was described by Vanderbilt ultrasoft pseudopotentials. Within the method used, the electronic wave functions were expanded in a plane wave basis set with the energy cut-off of 310~eV. The electrons 4$\textit{d}^{10}$5$\textit{s}^{2}$ for Cd, 5$\textit{s}^{2}$5$\textit{p}^{4}$ for Te and 4$\textit{s}^{2}$4$\textit{p}^{4}$ for Se atoms were taken as the valence ones. For DFT calculations of Cd${}_{16}$Se${}_{15}$Te solid state solution, the 2$\times$1$\times$2 supercell containing 32 atoms was created. The 2$\times$4$\times$2 Monkhorst-Pack mesh was used for the Brillouin zone (BZ) sampling~\cite{36}. The self-consistent convergence of the total energy was taken as 5.0$\times$10${}^{-7}$~eV/atom. The triclinic symmetry $\textit{P}$1 was kept during structure optimization of the crystal. Geometry optimization of the lattice parameters and atomic coordinates was performed using the Broyden-Fletcher-Goldfarb-Shanno (BFGS) minimization technique with the maximum ionic Hellmann-Feynman forces within 0.01 eV/\AA, the maximum ionic displacement within 5.0$\times$10${}^{-4}$~\AA, and the maximum stress within 0.02~GPa. These parameters are sufficiently small to lead to a well-converged total energy of the structures studied.

\section{Results and discussion}

\begin{figure}[!t]
\centerline{\includegraphics[width=10cm]{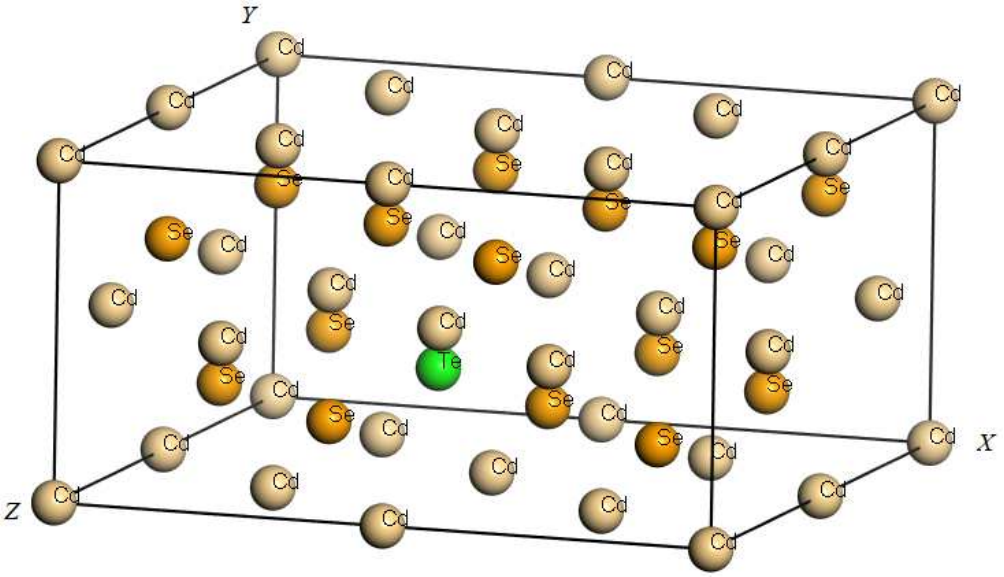}}
\caption{(Colour online) View of Cd${}_{16}$Se${}_{15}$Te crystal 2$\times$1$\times$2 supercell.}
\label{f1}
\end{figure}

The optimized lattice parameters of  Cd${}_{16}$Se${}_{15}$Te crystal (figure~\ref{f1}) are presented in table~\ref{tabular:1}.

\begin{table}[!t]
\caption{The calculated parameters of solid state solution Cd${}_{16}$Se${}_{15}$Te ($\textit{a}$ --- lattice parameters;  $\textit{E}_{g}$ --- band gap;  $\textit{E}_{\text{HH}}$ --- heavy hole band at G point;  $\textit{E}_{\text{LH}}$ --- light hole band at G point; $\Delta$ --- split-off energy at G point; $\textit{m}_{\text{HH}}^{\text{GF}}$, $\textit{m}_{\text{LH}}^{\text{GF}}$, $\textit{m}_{\text{SO}}^{\text{GF}}$, $\textit{m}_{\text{CB}}^{\text{GF}}$, $\mu$ --- heavy hole, light hole, spin-orbit, conduction
band effective mass and reduced mass in G point, respectively). Definitions of the symbols used are the relations~(\ref{eq1}).}
\label{tabular:1}
\begin{center}
	{\setlength{\extrarowheight}{4.5pt}
\vspace{0.3cm}
\begin{tabular}{|c|c|}
\hline
$\textit{a}$, nm & 0.6397 \\
\hline
$\textit{E}_{g}$, eV &	0.967 \\
\hline
$\textit{E}_{\text{HH}}$, meV & 0 \\
\hline
$\textit{E}_{\text{LH}}$, meV &	$-3.55$ \\
\hline
$\Delta$, meV &	13.74 \\
\hline
$\textit{m}_{\text{HH}}^{\text{GF}}$, $\textit{m}_{e}$ &	0.41 \\
\hline
$\textit{m}_{\text{LH}}^{\text{GF}}$, $\textit{m}_{e}$ &	0.29 \\
\hline
$\textit{m}_{\text{SO}}^{\text{GF}}$, $\textit{m}_{e}$ &	0.35 \\
\hline
$\textit{m}_{\text{CB}}^{\text{GF}}$, $\textit{m}_{e}$ &	0.16  \\
\hline
$\mu$, $\textit{m}_{e}$ &	0.12  \\
\hline
\end{tabular} }
\end{center}
\end{table}

\begin{figure}[!t]
\centerline{\includegraphics[width=9cm]{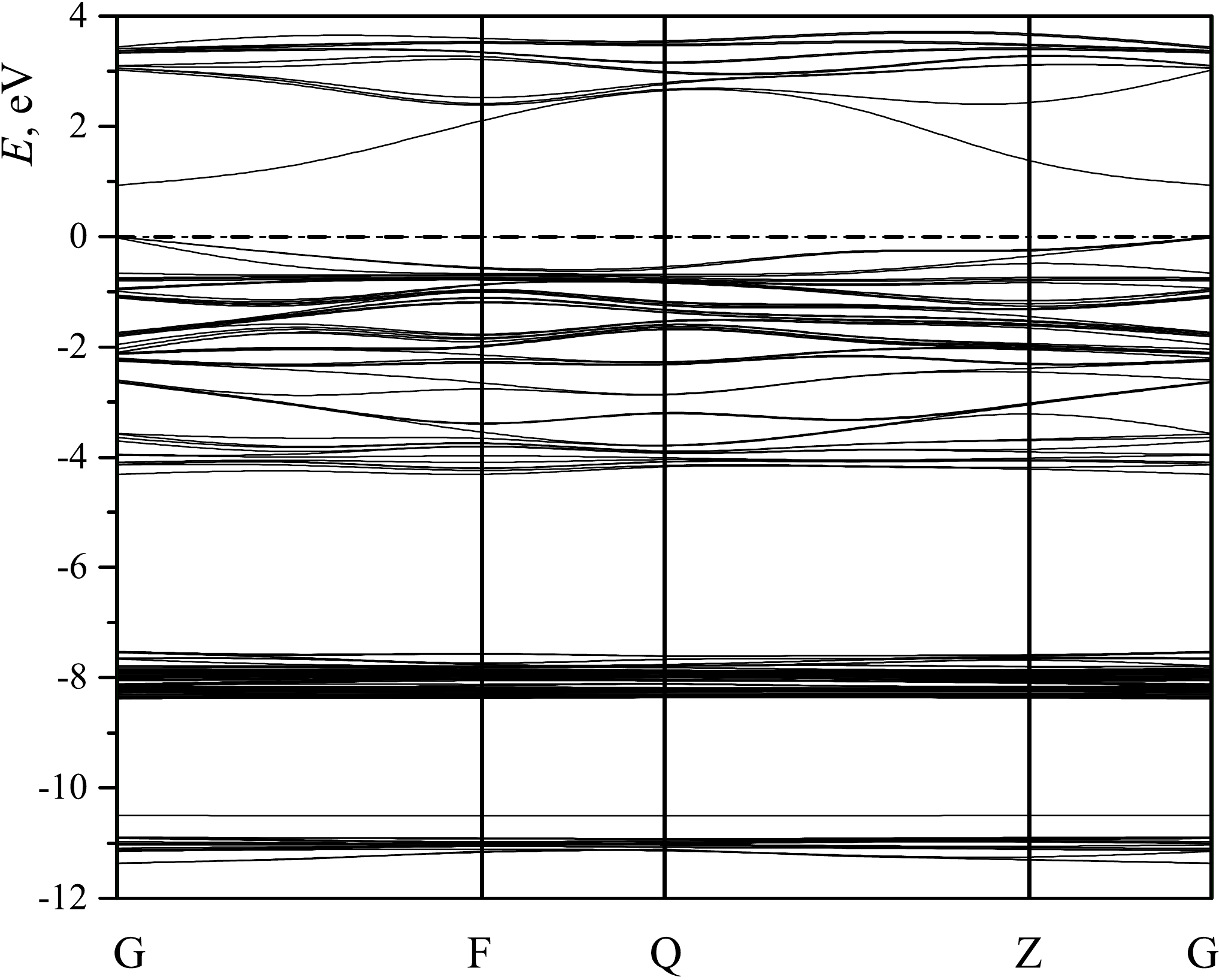}}
\caption{Energy band structure of Cd${}_{16}$Se${}_{15}$Te solid state solution for the points of Brillouin zone G(0,~0,~0), F(0,~0.5,~0), Q(0,~0.5,~0.5) and Z(0,~0,~0.5) (space group no.1, $\textit{P}$1).\label{f2}}
\end{figure}

The calculated band structure of Cd${}_{16}$Se${}_{15}$Te is presented in figure~\ref{f2}. Here, the Fermi level ${E}_{\text{F}}$ corresponds to the energy ${E}= 0$. The band gap of Cd${}_{16}$Se${}_{15}$Te is found to be of the direct type, ${E}_{g}^{\text{dir}}$ (figure~\ref{f2}). Thus, in this material, the indirect optical transitions, involving phonons, may occur at the photon energies $h\nu> {E}_{g}^{\text{dir}}$. This is similar to the cases realized in the corresponding binary compounds CdSe and CdTe~\cite{22, 25}.

The analysis of theoretical calculations of the energy band spectrum shows that the direct band gap $\textit{E}_{g}$ is localized at the center of BZ (at G point). The calculated value of the band gap $\textit{E}_{g}$ of Cd${}_{16}$Se${}_{15}$Te is estimated as $\textit{E}_{g}= 0.967$~eV, which is smaller than the experimental value for pure CdSe (1.68~eV~\cite{22}) and CdSe${}_{0.935}$Te${}_{0.065}$ (1.62~eV~\cite{10}). In this relation, it is well known that DFT-based calculations of semiconductors in the LDA and GGA levels of theory usually underestimate the band gap $\textit{E}_{g}$~\cite{25}. Thus, the fundamental optical absorption edge of Cd${}_{16}$Se${}_{15}$Te is formed by the direct inter-band electron transitions (the same situation is realized in other binary compounds of cadmium containing A${}^{\text{II}}$B${}^{\text{VI}}$ materials~\cite{22, 25, 37}). Using the experimental data of the band gap~\cite{10} we have obtained the value of ``scissor'' factor being equal to 0.653~eV (the ``scissor'' factor may be used for comparison of theoretical and experimental optical spectra in the range of electron excitations). The effective electron masses $\textit{m}^{*}$~(\ref{eq2}) calculated on the basis of the band structure obtained are presented in table~\ref{tabular:1}.

The interaction between the host matrix (CdSe) and the nearby placed tellurium atoms ($\textit{E}_{\text{Te}}$ and $\textit{E}_{\text{Te-SO}}$) causes the energy splitting into the heavy hole (HH) and light hole (LH) bands and the spin-orbit (SO) splitting of valence bands, while the conduction bands (CB) remain unaffected. The same situation is realized in GaSb${}_{1-\textit{x}}$Bi${}_{\textit{x}}$ solid state solution~\cite{38}.

In addition, a clear difference of the band dispersions $\textit{E}$($\textit{k}$) between the valence and conduction bands is observed. The top valence bands are more flat than the bottom conduction bands (figure~\ref{f2}), which is caused by the larger valence hole effective masses in comparison to the effective masses of the conduction electrons~\cite{22, 39}. The electron band dispersions $E(k)$ near the G-point are defined as~\cite{31}:

\begin{equation}
\textit{E}_{\text{SO}}= -\Delta\textit{E}_{0} - \frac{\hslash^{2}\textit{k}^{2}}{2\textit{m}_{\text{SO}}}, \quad  \textit{E}_{\text{LH}}= \frac{\hslash^{2}\textit{k}^{2}}{2\textit{m}_{\text{LH}}}, \quad \textit{E}_{\text{HH}}= \frac{\hslash^{2}\textit{k}^{2}}{2\textit{m}_{\text{HH}}}, \quad \textit{E}_{e}= \textit{E}_{g} - \frac{\hslash^{2}\textit{k}^{2}}{2\textit{m}_{e}},
\label{eq1}
\end{equation}
where $\Delta\textit{E}_{0}$ is the spin-orbit splitting, $\textit{E}_{g}$ is the band gap, $\textit{m}_{i}$ stands for the different associated masses, and the $\textit{k}$ --- wave vector is evaluated in such a way that we are in the parabolic regime of the electronic band, usually around 0.5\% of the full wave vector magnitude in each direction. Different effective masses are obtained from the relation:

\begin{equation}
\frac{1}{\textit{m}^{*}}= \frac{4\piup^{2}}{h^{2}}\frac{\text{d}^{2}\textit{E}(\textit{k})}{\text{d}\textit{k}^{2}},
\label{eq2}
\end{equation}
where $\textit{h}$ is the Planck constant, $\textit{E}$($\textit{k}$) is the dependence of the band energy $\textit{E}$ on the electron wave vector~$\textit{k}$.

The calculated effective masses are presented in table~\ref{tabular:1}. Since CdSe, CdS, CdTe are important materials for solar cells and electronic devices, their electronic conductivity parameters are very important for practical application. The comparison of our calculated data of effective mass for Cd${}_{16}$Se${}_{15}$Te with the binary compounds (CdTe and CdSe) shows practically the same value which is observed for CdSe (0.13$\textit{m}_{e}$~\cite{40, 41}). The comparison of our calculated values compared to literature data is listed in table~\ref{tabular:2}. The similar SO effective mass for CdTe crystal and Cd${}_{16}$Se${}_{15}$Te solid state solution indicates that the spin-orbit splitting is associated with the Te energy levels located nearby.

\begin{figure}[!t]
\centerline{\includegraphics[width=10cm]{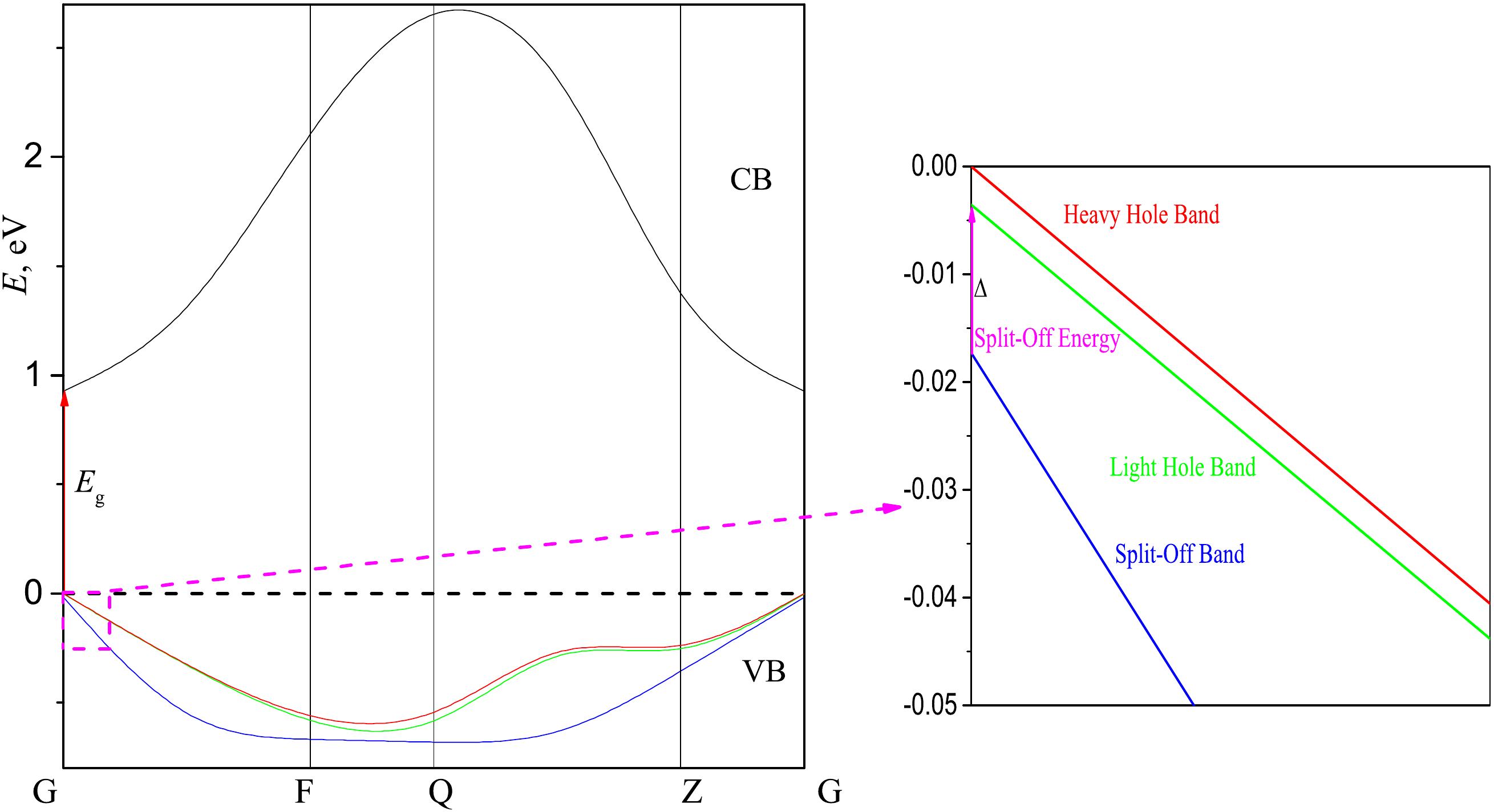}}
\caption{(Colour online) Part of the energy band diagram of Cd${}_{16}$Se${}_{15}$Te solid state solution.\label{f3}}
\end{figure}

\begin{table}[!b]
\caption{Effective masses for CdSe and CdTe in [100] and [111] directions and for Cd${}_{16}$Se${}_{15}$Te in [010] direction. All the masses are given in units of the free electron mass $\textit{m}_{e}$.}
\label{tabular:2}
\vspace{0.3cm}
\begin{center}
\begin{tabular}{|c|c|c|c|c|c|}
\hline
Compound & Direction & $\textit{m}_{\text{CB}}/\textit{m}_{e}$ &$\textit{m}_{\text{HH}}/\textit{m}_{e}$& $\textit{m}_{\text{LH}}/\textit{m}_{e}$&$\textit{m}_{\text{SO}}/\textit{m}_{e}$\\
\hline
CdTe~\cite{42, 43} &	[100]	&0.110&	0.600&	0.180&	0.350\\
\hline
 & [111]& 	0.09& 	0.40& 	-& 	- \\
\hline
CdSe~\cite{42, 44} &	[100]&	0.119&	0.820&	0.262&	-\\
\hline
 &	[111]	&0.11&	0.45&	-	&- \\
\hline
Cd${}_{16}$Se${}_{15}$Te &	[010]&0.16&	0.41&	0.29	&0.35 \\
\hline
\end{tabular}
\end{center}
\end{table}

The refractive index $\textit{n}$ is a closely related value to the electronic properties and band structure of the material. Many empirical relations between refractive index $\textit{n}$ and band gap $\textit{E}_{g}$ were proposed earlier~\cite{45, 46, 47, 48, 49, 50, 51, 52}. According to the Tripathy relation~\cite{53}, the refractive index of a semiconductor with the band gap $\textit{E}_{g}$ is given by:

\begin{equation}
\textit{n}= \textit{n}_{0}\left[1+\alpha \re^{-\beta\textit{E}_{g}}\right],
\label{eq3}
\end{equation}
where, $\textit{n}_{0}= 1.73$, $\alpha= 1.9017$ and $\beta= 0.539$~eV${}^{-1}$ are constant parameters for certain ranges of temperature and pressure. The energy region, in which this relation satisfactorily fits the corresponding experimental dependences, is within the range $0<\textit{E}_{g}<5$~eV. Some other empirical relations available in literature are Moss~\cite{45}, Ravindra~\cite{47} and Herve-Vandamme (HV) ones~\cite{48, 49} (see equation~\ref{eq4}, \ref{eq5} and~\ref{eq6}, respectively).
\begin{equation}
\textit{n}^{4}\textit{E}= 95 \text{ eV},
\label{eq4}
\end{equation}
\begin{equation}
\textit{n}= 4.084 - 0.62\textit{E}_{g},
\label{eq5}
\end{equation}
\begin{equation}
\textit{n}^{2}= 1 + \left(\frac{A}{\textit{E}_{g} + B}\right)^{2},
\label{eq6}
\end{equation}
where $\textit{A}$ is the hydrogen ionization energy 13.6~eV and $\textit{B}= 3.47$~eV is a constant parameter assumed to be the difference between UV resonance energy and band gap $\textit{E}_{g}$. The calculated values of $\textit{n}$ using various models of interest are listed in table~\ref{tabular:3} (the calculated band gap $\textit{E}_{g}$ of Cd${}_{16}$Se${}_{15}$Te includes the “scissor” factor).

Based on the values of $\textit{n}$ calculated from different relations, the high-frequency dielectric constant $\varepsilon_{\infty}$= $\textit{n}^{2}$ and the exciton binding energy $\textit{E}_{b}$ are determined for Cd${}_{16}$Se${}_{15}$Te solid state solution (table~\ref{tabular:3}). The exciton binding energy $\textit{E}_{b}$ is calculated using the Bohr model~(\ref{eq7})

\begin{equation}
\textit{E}_{b}= 13.6 \text{ eV } \frac{\mu}{\varepsilon_{\infty}^{2}},
\label{eq7}
\end{equation}
where $\mu$ is the reduced mass (for calculation of $\mu$ we used the electron and heavy-hole masses) and $\varepsilon_{\infty}$  is the high-frequency dielectric constant.

\begin{table}[!t]
\caption{Refractive index $\textit{n}$, dielectric constant $\varepsilon_{\infty}$ and exciton binding energy $\textit{E}_{b}$ for Cd${}_{16}$Se${}_{15}$Te solid state solution.}
\label{tabular:3}
\vspace{0.3cm}
\begin{center}
\begin{tabular}{|c|c|c|c|c|c|}
\hline
Model&	Tripathy relation&	Moss relation&	Ravindra relation&	HV relation&	Exp. for CdSe\\
\hline
$\textit{n}$&	3.10&	2.77&	3.08&	2.85&	~2.82~\cite{54}\\
\hline
$\varepsilon_{\infty}$&	9.61	&7.67	&9.48	&8.12&	5.72~\cite{55}\\
\hline
$\textit{E}_{b}$, meV &	17.67 &	27.74	 &18.16 &	24.75 &	16$\pm$1.5~\cite{56}\\
\hline
\end{tabular}
\end{center}
\end{table}

The comparison of different calculation methods of refractive index with experimental results shows the best correlation with Tripathy model (table~\ref{tabular:3}). The calculated value of exciton binding energy $\textit{E}_{b}$ is in good agreement with the experimental results for cubic phase of CdSe~\cite{56}. The calculated value of $\textit{E}_{b}$ (table~\ref{tabular:3}) is also close to the similar value for the bulk silicon ($\textit{E}_{b}(\text{Si})=15$~meV~\cite{57}).

Elastic properties play an important role in providing valuable information on the bonding characteristics between adjacent atomic planes. They are capable of determining how the material undergoes the stress deformation, and then recovers and returns to its original shape after stress cessation. Furthermore, these properties play an important role in providing valuable information on structural stability, anisotropic factors, Debye temperature, phonon spectra and specific heat. All this information is usually defined by the elastic constants $\textit{C}_{ij}$~\cite{58, 59}. The calculation method used allows the calculation of the total energy $\textit{E}$ for arbitrary crystal structures.

One can deform the calculated equilibrium crystal structure, determine the total energy of the strained crystal $\textit{E}$ and use the obtained results to estimate the elastic constants. The elastic constants are proportional to the second-order coefficient in the polynomial expansion of the total energy $\textit{E}$ as a function of the strain parameter $\delta$. Only small deformations, which did not exceed the crystal elasticity limit, were taken into consideration in calculations. Knowing the total crystal energy $\textit{E}$ and its variation by the strain~$\delta$, one can determine nine elastic constants from the following equations:

\begin{eqnarray}
&&E(V,\delta)=E(V_{0},0)+V_{0} (\tau_{1}\delta+C_{11}\delta^{2}/2), \nonumber\\
&&E(V,\delta)=E(V_{0},0)+V_{0} (\tau_{2}\delta+C_{22}\delta^{2}/2), \nonumber\\
&&E(V,\delta)=E(V_{0},0)+V_{0} (\tau_{3}\delta+C_{33}\delta^{2}/2), \nonumber\\
&&E(V,\delta)=E(V_{0},0)+V_{0} (2\tau_{4}\delta+2C_{44}\delta^{2}), \nonumber\\
&&E(V,\delta)=E(V_{0},0)+V_{0} (2\tau_{5}\delta+2C_{55}\delta^{2}), \nonumber\\
&&E(V,\delta)=E(V_{0},0)+V_{0} (2\tau_{6}\delta+2C_{66}\delta^{2}), \nonumber\\
&&E(V,\delta)=E(V_{0},0)+V_{0} \left[(\tau_{1}-\tau_{2})\delta+ \frac{(C_{11} + C_{22} -2C_{12})\delta^{2}}{2}\right], \nonumber\\
&&E(V,\delta)=E(V_{0},0)+V_{0} \left[(\tau_{1}-\tau_{3})\delta+ \frac{(C_{11} + C_{33} -2C_{13})\delta^{2}}{2}\right], \nonumber\\
&&E(V,\delta)=E(V_{0},0)+V_{0} \left[(\tau_{2}-\tau_{3})\delta+ \frac{(C_{22} + C_{33} -2C_{23})\delta^{2}}{2}\right]. 
\label{eq8}
\end{eqnarray}

In the equation~(\ref{eq8}), $\textit{V}$ is the supercell volume. Here, the elastic constants $\textit{C}_{12}$, $\textit{C}_{13}$, and $\textit{C}_{23}$ were determined as linear combinations of the already obtained constants of $\textit{C}_{11}$, $\textit{C}_{22}$, and $\textit{C}_{33}$. The calculated elastic constants $\textit{C}_{ij}$ of Cd${}_{16}$Se${}_{15}$Te are presented in table~\ref{tabular:4}.

\begin{table}[th]
\caption{The calculated elastic constants of Cd${}_{16}$Se${}_{15}$Te solid state solution.}
\label{tabular:4}
\begin{center}
\begin{tabular}{|c|c|}
\hline
Symbol of elastic constant $\textit{C}_{ij}$&	Value of elastic constant $\textit{C}_{ij}$ (GPa)\\
\hline
$\textit{C}_{11}$&	41.19$\pm$4.64\\
\hline
$\textit{C}_{22}$&	48.55$\pm$3.01\\
\hline
$\textit{C}_{33}$&	42.12$\pm$4.60\\
\hline
$\textit{C}_{44}$ &	36.90$\pm$2.87\\
\hline
$\textit{C}_{55}$&	31.12$\pm$5.04\\
\hline
$\textit{C}_{66}$&	36.89$\pm$2.87\\
\hline
$\textit{C}_{12}$&	41.36$\pm$4.60\\
\hline
$\textit{C}_{13}$ &	55.41$\pm$1.28\\
\hline
$\textit{C}_{23}$ &	41.28$\pm$4.58\\
\hline
\end{tabular}
\end{center}
\end{table}

The theoretical polycrystalline elastic modulus of Cd${}_{16}$Se${}_{15}$Te solid state solution can be determined by two approximation methods, namely the Voigt and the Reuss methods~\cite{60}. Voigt assumes the uniform strain throughout the polycrystalline aggregate and Reuss assumes a uniform stress. The bulk modulus~$\textit{B}$, Young’s modulus~$\textit{E}$, shear modulus $\textit{G}$ and Poisson’s ratio $\sigma$ are calculated directly by the Voigt-Reuss-Hill (VRH) method~\cite{61} (table~\ref{tabular:5}).

\begin{table}[!t]
\caption{The calculated values of shear modulus ($\textit{G}$), bulk modulus ($\textit{B}$), Young’s modulus ($\textit{E}$), and Poisson’s ratio ($\sigma$) of Cd${}_{16}$Se${}_{15}$Te solid state solution.}
\label{tabular:5}
\begin{center}
\vspace{0.3cm}
\begin{tabular}{|c|c|c|c|}
\hline
&	Voigt	&Reuss	&Hill\\
\hline
$\textit{B}$, GPa&	45.44	&44.98	&45.21\\
\hline
$\textit{G}$, GPa&	20.64&	28.97	&27.80\\
\hline
$\textit{B/G}$	&2.20&	1.55	&1.62\\
\hline
$\textit{E}$, GPa&	53.78	&71.55	&69.21\\
\hline
Poisson ratio ($\sigma$)&	0.30&	0.23	&0.25\\
\hline
\end{tabular}
\end{center}
\end{table}

According to the elastic criteria, the material is brittle (ductile) if the $\textit{B/G}$ ratio is less (greater) than 1.75. The calculated values $\textit{B/G}$ of Cd${}_{16}$Se${}_{15}$Te are smaller than 1.75~\cite{62} when they are obtained using the Reuss and Hill methods. Hence, the material studied should probably behave in a brittle manner. The Poisson's ratio of a stable, isotropic, linear elastic material should be between $-1.0$ and $+0.5$ because of the requirement for Young's modulus, and the shear modulus and bulk modulus should have positive values~\cite{63}. According to Frantsevich rule~\cite{64}, the critical value of Poisson ratio of a material is $1/3$~\cite{65}. The value of Poisson’s ratio $\sigma$, responsible for ductile ($\sigma> 1/3$) or brittle ($\sigma< 1/3$) character, corresponds in our case to the brittle one ($\sigma<1/3$). The value of the Poisson’s ratio is indicative of the degree of directionality of the covalent bonds. This value is relatively small ($\sigma= 0.1$) for the covalent materials and relatively large ($\sigma= 0.25$) for the ionic ones. The calculated Poisson’s ratio $\sigma$ of Cd${}_{16}$Se${}_{15}$Te lies in the range of 0.23--0.30.

Knowing the elastic constants $\textit{C}_{ij}$ of a material one can calculate the corresponding acoustic velocities in certain directions. The values of acoustic velocity in different directions of Cd${}_{16}$Se${}_{15}$Te crystal were calculated using the respective relations~\cite{66}:

\begin{equation}
\vartheta_{l}= \sqrt{\frac{3\textit{B}+4\textit{G}}{3\rho}},
\label{eq9}
\end{equation}

\begin{equation}
\vartheta_{t}= \sqrt{\frac{\textit{G}}{\rho}},
\label{eq10}
\end{equation}
where $\rho$ is the density of a crystal. The calculated value of density $\rho$ is 4.94~g~cm${}^{-3}$. This value is in good agreement with the experimental value for pure CdSe~\cite{67}. The calculated acoustic velocities in different directions are presented in table~\ref{tabular:6}.

\begin{table}[th]
\caption{The calculated acoustic velocities in different directions of Cd${}_{16}$Se${}_{15}$Te solid state solution.}
\label{tabular:6}
\begin{center}
\begin{tabular}{|c|c|c|c|}
\hline
&	Voigt	&Reuss	&Hill\\
\hline
\hline
$\vartheta_{l}$, m s${}^{-1}$&	3843.1	&4113.9	&4081.1\\
\hline
$\vartheta_{t}$, m s${}^{-1}$&	2044.1&	2421.7	&2372.2\\
\hline
$\theta_{\text{D}}$, K&	273.81	&303.97	&300.14\\
\hline
\end{tabular}
\end{center}
\end{table}

The Debye temperature $\theta_{\text{D}}$ is one of the most important parameters which determines the thermal properties of a material. The Debye temperature can be defined in terms of the mean acoustic velocity and gives explicit information on the lattice vibrations. This is the highest temperature that corresponded to the highest frequency normal vibration $\nu_\text{D}$ ($\theta_{\text{D}}$= $h\nu_\text{D}/k_{\text{B}}$, where $\textit{k}_{\text{B}}= 1.380658\times 10^{-23}$~J~K${}^{-1}$). At relatively low temperatures, vibrational excitations arise mainly due to acoustic oscillations. Therefore, the value of $\theta_{\text{D}}$, calculated from the elastic constants, is the same as that determined by specific heat measurements at relatively low temperatures.

Using the mean acoustic velocity, the Debye temperature is calculated from equation~(\ref{eq12})

\begin{equation}
	\theta_{\text{D}}= \frac{\hslash}{\textit{k}_{\text{B}}}\left(\frac{6\piup^{2}\textit{N}}{\textit{V}_{0}}\right)^{1/3} \overline{\vartheta},
	\label{eq12}
\end{equation}
where $\textit{N}$ is the number of atoms in the supercell of Cd${}_{16}$Se${}_{15}$Te, $\textit{V}_{0}$ is the supercell volume. The obtained value of Debye temperature is in good correlation with other known $\theta_{\text{D}}$ values for binary compounds ($\theta_{\text{D}}= 295.6$~K for CdTe and $\theta_{\text{D}}= 317.6$~K for CdSe obtained at the temperature 298~K~\cite{68}).

\section{Conclusion}

In this work, the electron band structure, effective mass, optical parameters, acoustic velocity, Debye temperatures and elastic properties of Cd${}_{16}$Se${}_{15}$Te solid state solution are estimated by first-principles calculations. The calculations were performed within the generalized gradient approximation (GGA) with the Perdew-Burke-Ernzerhof (PBESOL) exchange-and-correlation functional. Using Voigt-Reuss-Hill approximation, we calculated and discussed the ideal polycrystalline aggregates bulk modulus, shear modulus, Young’s modulus, and Poisson’s ratio. The obtained values are in good agreement with experimental data and correlate with the values observed for binary compounds CdSe and CdTe.

\section{Acknowledgements}

The work was supported by the Project 0119U002247 of Ministry of Education and Science of Ukraine. Computer simulations are performed using the CASTEP code at the ICM of the Warsaw University (the Project GB81-13) and at the WCSS of the Wroclaw University of Technology (the~Project~053).

\ukrainianpart

\title{Електронна структура та пружні постійні твердого розчину заміщення Cd${}_{16}$Se${}_{15}$Te: дослідження з перших принципів}
\author{А.~І. Кашуба\refaddr{label1}, Б. Андрієвський\refaddr{label2}, Г.~А. Ільчук\refaddr{label1}, Р.~Ю. Петрусь\refaddr{label1}, Т.~С. Малий\refaddr{label3}, \\ І.~В. Семків\refaddr{label1}}
\addresses{
\addr{label1} Національний університет ``Львівська політехніка'', вул. С. Бандери, 12, 79013 Львiв, Україна
\addr{label2} Кошалінський технологічний університет, вул. Снядечких, 2, 75-453 Кошалін, Польща
\addr{label3} Львівський національний університет імені Івана Франка, вул. Кирила і Мефодія, 8, 79005 Львів, Україна
}

\makeukrtitle

\begin{abstract}
\tolerance=3000%
Подаються результати дослідження з перших принципів електронного енергетичного спектру та пружних постійних твердого розчину заміщення Cd${}_{16}$Se${}_{15}$Te. Структура досліджуваного зразка була побудована на основі ``батьківського'' бінарного з'єднання CdSe, який кристалізується в кубічній фазі. На основі електронної енергетичної структури було обчислено ефективні маси електрона, важкої дірки, легкої дірки, ефективну масу для енергетичного рівня спін-орбітального з'єднання та зведену масу в G точці. Крім того, подано результати розрахунку енергії зв'язку екситона, показника заломлення та діелектричної постійної високої частоти. Модуль Юнга, модуль зсуву, об'ємний модуль пружності та відношення Пуассона розраховані з перших принципів. За результатами пружних постійних було отримано значення швидкості звуку та температури Дебая.
\keywords твердий розчин заміщення, електронна енергетична структура, ефективна маса, пружні постійні, температура Дебая

\end{abstract}
\lastpage
\end{document}